\documentclass[11pt]{article}

 \usepackage{amsmath}
 \usepackage{graphicx}
 \usepackage{amssymb}
 \usepackage{amsfonts}
 \usepackage{latexsym}

\renewcommand{\baselinestretch}{1.1}
\renewcommand{\vec}[1]{{\bf #1}}

\def\beq{\begin{eqnarray}}
\def\eeq{\end{eqnarray}}

\def\al{\alpha}
\def\be{\beta}

\def\ga{\gamma}
\def\de{\delta}

\def\la{\lambda}
\def\na{\nabla}

\def\si{\sigma}

\def\Ga{\Gamma}

\def\Si{\Sigma}

\begin{document}
\begin{center}
{\large\bf The Exact Foldy-Wouthuysen Transformation for a Dirac Theory Revisited}
\vskip 6mm
\textbf{Bruno Gon\c calves\footnote{
E-mail: bruno.goncalves@ifsudestemg.edu.br},
\
M\' ario M. Dias J\' unior\footnote{
E-mail: mariodiasjunior@ice.ufjf.br}
\ and \
Baltazar J. Ribeiro\footnote{
E-mail: baltazarjonas@cefetmg.br}
}
\vskip 4mm
$^{1}$Instituto Federal de Educa\c c\~ ao, Ci\^ encia e Tecnologia Sudeste de Minas Gerais \\
\centering{IF Sudeste MG, 36080-001, Juiz de Fora - MG, Brazil} \vspace{0.4cm}\\

$^{2}$ Departamento de F\' isica, ICE, Universidade Federal de Juiz de Fora,\\
UFJF, 36036-330, Juiz de Fora - MG, Brazil \vspace{0.4cm}\\

$^{3}$Centro Federal de Educa\c c\~ ao Tecnol\' ogica de Minas Gerais\\
CEFET-MG, 36700-000, Leopoldina - MG, Brazil

\end{center}
\vskip 4mm
\begin{quotation}

\begin{abstract}
The Exact Foldy-Wouthuysen transformation (EFWT) method is generalized here. 
In principle, it is not possible to construct the EFWT to any
Hamiltonian. The transformation conditions are the same but the involution
operator has a new form. We took a particular example and constructed
explicitly the new involution operator that allows one to perform the
transformation. We treat the case of the Hamiltonian with 160 possible
CPT-Lorentz breaking terms, using this new technique. The transformation
was performed and physics analysis of the equations of motion is shown. 

\end{abstract}


\vspace{0.2cm}
\end{quotation}
\vskip 4mm

\section{Introduction}

\hspace{0.65cm}

The study of the possible candidates to break CPT-Lorentz symmetry is very important nowadays \cite{kost2018}. There are a large study been developed during the last ten years that shows the possible experiments that could give the more prominent physical effect to measure one of these fields \cite{tables}. Until now, non of them was directly observed. The most prominent theoretical approaches that consider these cases are based on indirect physical effects, as it is shown in \cite{colladay, kostrev}.  In other words, the search for this manifestations starts with an action that considers at least two independent fields as one can see on the recent papers \cite{ding}. For the non-relativistic 
scenario the the results are well established in \cite{SPS, BBSh, RySh, Lammerzahl}, for torsion field, for example. It is very interesting to see \cite{shapirorep} that torsion field could be generated from the symmetry breaking. Some recent theoretical studies have been developed 
with the same phenomenological background \cite{Alex, CFMS, ACF, KaiMa}.

Another possible phenomenological approach to this problem can be constructed step by step by searching for new terms in the Hamiltonian that describes this situation. Thinking this way, it makes sense the appearance of some terms in the equations of motion that could give a mix between an external known field with sufficient enough big amplitude to compensate the fact that the CPT-Lorentz terms have small amplitudes. 

The idea is the same shown in \cite{bruno1}, where the strong magnetic field could, in principle, change the trajectory of the Dirac particle that interacts with gravitational waves. It is important to take into account the corrections, made with canonical FW,  to these results that were shown in \cite{bruno1COR}. In \cite{IvPitWell}, the massive linearized gravity was studied and some possible experiments that could measure indirect effects of gravitational waves on Dirac fermions were indicated.  However, solving the Dirac equation for the general case is not a simple procedure \cite{kostebase}. It is well known in literature that working with the EFWT is a more prominent 
approach to interpret a Dirac Hamiltonian than the canonical transformation \cite{obukhov}. But this is true not only for the fact that it can give us new terms, but it is a faster and more economic (in terms of algebraic calculation) procedure \cite{bruno1, MurRaya, nikitin, bruno2}. One can see this transformation as a generalization of the usual FWT. 

Let us perform a comparison on the two procedures. It is possible to see that 
in the usual FWT the multiplication on each step (on each order on $1/m$) 
by the term that makes the Hamiltonian even, generates a maximum of $1+2n$ even 
terms, where $n$ represents the number of terms of the previous Hamiltonian (see, for example, pages 48-51 in \cite{bjorken}). The maximum number of terms in the $nth$-Hamiltonian is straightforward obtained by the fact that this is an expansion 
in power series of an operator. The factor $2$ on $1+2n$ expression is obtained in case when it does not commute with all original terms. 

On the other hand, the EFWT impose the multiplication of all terms of 
the Hamiltonian by themselves. Analogous arguments give us the maximum of 
$1+2n^2$ on the expanded Hamiltonian. If the parameter of expansion here 
is also taken to be $1/m$, one can see that the possibility of having new 
terms in comparison with the usual method is greater. In many particular 
known cases \cite{nikitin, eriksen, case}, the anti-commutators on both 
cases are such that the results are the same! But it is not the general 
case. This was explicitly shown on \cite{obukhov}. In this paper we show 
another case where it happens. 

In \cite{silenko2013} the author performs in a very didactic way the formal comparison between the two methods. He also described which is the most efficient method for each possible applications. The explicit calculations are performed in the series of three works where the generality for the exact procedure becomes evident \cite{silenko1, silenko2, silenko3}. 

On \cite{bruno3, FFP14, OP}, the authors worked on a series of papers in which the EFWT conditions were not satisfied. In these articles, the study of the CPT-Lorentz violating terms was used as a background to this transformation. It is possible to see in \cite{bmb} the diagonalized Hamiltonian for all the possible terms that allows this procedure. 

Using the result of \cite{OP}, we developed an algorithm to construct a generalized involution operator for the EFWT. We show a method to construct the explicit form of the operator that allows the Hamiltonian to be diagonalized. In some sense, the logic here is inverse: we do not test if it is possible to perform the EFWT but we search for the operator that gives us this possibility. 

By showing the explicit analytic form of this operator, the EFWT usual algorithm can be applied to the initial Hamiltonian. We constructed the general operator and the complete case of CPT-Lorentz interacting with Dirac field \cite{kostebase} is studied here using the EFWT technique. We also compare the result with the usual transformation and two new terms show up.

%


\section{The complete Hamiltonian for a Dirac Theory with\\ CPT-Lorentz invariance violation}

\hspace{0.65cm}In Ref. \cite{bruno2}, the authors present a table that specifies 
the $80$ cases of CPT and Lorentz 
violating terms in the modified Dirac equation. 
A complete study of the EFWT, taking into account these $80$ cases, is presented in Ref. \cite{bmb}. 

However, it is worth mentioning that, in Refs \cite{bruno2,bmb}, a sort of terms were not considered. 
In order to perform the EFWT 
study of the complete set of cases, it is necessary that the Hamiltonian admits the involution operator \cite{obukhov, nikitin, eriksen, case, bmb}.
In this work, we present a new table corresponding to all the CPT-Lorentz breaking terms. 
The main point is the search for an involution operator $J$, which satisfies 
the anti-commutation relation, 
\begin{equation}
JH + HJ = 0, \label{atrfg} 
\end{equation}
for the complete set of terms, presented in the table.

\begin{table}[htb]
   \centering
   \setlength{\arrayrulewidth}{1.5\arrayrulewidth}  
   \setlength{\belowcaptionskip}{10pt}  
   \caption{\it Interaction coefficients}
   \resizebox{\textwidth}{!}{%
   \begin{tabular}{|c|c|c|c|c|c|c|c|c|c|}
      \hline
& $m$ & $a_l$ & $b_0$ & $H^{lj}$ & & $m_5$& $b_l$& $a_0$& $H^{0 \mu}$ \\
& $P^*_\nu \, e^{\nu}$ & $P^*_\nu \, c^{l \nu}$ & $P^*_\nu \, d^{0\nu}$
& $P^*_\nu \, g^{lj\nu}$ & & $P^*_\nu \, f^\nu $& $P^*_\nu \, d^{l\nu}$& $P^*_\nu \, c^{0\nu}$
& $P^*_\nu \, g^{0\mu\nu}$ \\

& & $\overline{P}_l$ & & & & & & $\overline{P}_0$& \\
\hline \hline
$\ga^0$ & $1$ & $\ga^l$& $-\ga^0\ga^5$& $\frac{1}{2}\sigma^{lj}$
& & \boldmath$i\ga_5$ & \boldmath$\ga_5 \ga_l$ & \boldmath$ \ga_0 $&\boldmath $\frac{1}{2} \sigma^{0\mu}$\\
      \hline
$c^{00}$ & $-\ga^0$ & $-\al^l$& $\ga^5$& $-\frac{1}{2}\ga^0\sigma^{lj} $
& &\boldmath$-i\ga_0\ga_5$ &\boldmath$\ga_5\al_l$ &\boldmath $-1$ &\boldmath $-\frac{1}{2}\ga^0\sigma^{0\mu}$\\
      \hline
$f^0$ & $i\ga^5$ & $i\ga^5\ga^l$& $i\ga^0$& $\frac{i}{2}\ga^5 \sigma^{lj}$
& &\boldmath $-1$&\boldmath $i\ga^l$&\boldmath $i\ga^5 \ga^0$&\boldmath$\frac{i}{2}\ga^5 \sigma^{0\mu}$\\
      \hline
$d^{i0}$ & $-i\ga^i\ga^5$&  $i\ga^i\ga^5\ga^l$& $-\al^i$&
$-\frac{1}{2}\ga^i\ga^5 \sigma^{lj}$& & \boldmath$i\ga^i$ &\boldmath $\ga^i\ga^l$& \boldmath$\ga^5\al^i$& \boldmath$\frac{1}{2}\ga^i\ga^5 \sigma^{0\mu}$\\
      \hline
$g^{i00}$ & $-\frac{i}{2}\al^i$ & $-\frac{i}{2}\al^i\ga^l$& $-\frac{i}{2}\ga^i\ga^5$& $-\frac{1}{4}\al^i \sigma^{lj}$
& & \boldmath$\frac{1}{2}\al^i \ga_5$ & \boldmath$-\frac{i}{2}\al^i \ga_5 \ga_l$& \boldmath$-\frac{i}{2}\al^i \ga_0$ & \boldmath$-\frac{i}{4}\al^i \sigma^{0\mu}$ \\
      \hline 
& & & & & & & & & \\
			\hline 
$d^{00}$ &\boldmath${-\ga^0\ga^5}$ & \boldmath$\ga^5\al^l$&\boldmath$-1$ & \boldmath$\frac{1}{2} \ga_5 \ga_0 \sigma^{lj}$& & \bf $-i\ga^0$& $-\al^l$& $-\ga^5$
& $-\frac{1}{2}\sigma^{0 \mu}\ga^0\ga^5$\\
      \hline
$e^0$ &\boldmath$-1$ &\boldmath$-\ga^l$ &\boldmath$-\ga^5\ga^0$ & \boldmath$-\frac{1}{2}\sigma^{lj}$ & & $-i\ga^5$& $-\ga^5\ga^l$& $-\ga^0$
& $-\frac{1}{2}\sigma^{0 \mu}$\\
      \hline
$c^{i0}$ & \boldmath$ \ga^i$ &\boldmath$ \ga^i \ga^l$ & \boldmath$\ga^5 \al^i$ & \boldmath$\frac{1}{2}\ga^i\sigma^{lj}$& & $i\ga^i\ga^5$& $-i\ga^i\ga^5\ga^l$& $-\al^i$
& $\frac{1}{2}\ga^i\sigma^{0 \mu}$\\
      \hline
$g^{ik0}$ & \boldmath$-\frac{1}{2}\sigma^{ik}$ &\boldmath$-\frac{1}{2}\sigma^{ik}\ga^l$ &\boldmath$\frac{1}{2}\sigma^{ik}\ga^0\ga^5$ &  \boldmath$-\frac{1}{4}\sigma^{ik}\sigma^{lj}$& & $-\frac{i}{2}\sigma^{ik}\ga^5$
& $-\frac{1}{2}\sigma^{ik}\ga^5\ga^l$ & $-\frac{1}{2}\sigma^{ik}\ga^0$
& $-\frac{1}{4}\sigma^{ij}\sigma^{0 \mu}$\\
      \hline
   \end{tabular}
	}
\end{table}

The quantities $a_{\mu}$, $b_{\mu}$, $m_{5}$, $c^{\mu\nu}$, $d^{\mu\nu}$, $e^{\mu}$, $f^{\mu}$, $g^{\mu\nu\la}$ and $H_{\mu\nu}$ 
represent the CPT-Lorentz violating parameters \cite{kostelecky2,jackiw2,pickering}. 
We adopt notations as described in \cite{bjorken} 
for Dirac matrices and the 
useful notations for $P_i$, used in \cite{bruno2}.

The terms highlighted in boldface, correspond 
to the empty spaces, in the table presented in \cite{bruno2}. These terms do not obey the 
anti-commutation relation (\ref{atrfg}), if one takes into account the following form of the involution operator 
\begin{equation}
J=i\gamma^{5}\gamma^{0}\,.\label{oldj}
\end{equation}
The set of terms that obey relation (\ref{atrfg}), 
considering the involution operator (\ref{oldj}), are presented in 
Ref \cite{bruno2}.
%
From now on, we shall call the quantities in boldface as new terms and the quantities that are not in boldface, old terms.

In order to understand how the Hamiltonian can be obtained, directly from the table, let us 
present a simple example. The rule is based on the product of the line terms by the terms in the rows. 
We shall consider, for instance, the first line times the first 
row: $\ga^0 \times 1 \times m=\ga^0 m$. We got, in this case, the free Dirac equation term, which is the most trivial one. 

Let us consider another example. The product of the sixth line by the first row. The terms inside the table must also 
taken into account. Such a multiplying gives two terms 
\begin{equation}
d^{00}\times(-\ga^0\ga^5)\times m= -m d^{00} \ga^0\ga^5   
\hspace{0.5cm}\mbox{and}\hspace{0.5cm}
d^{00}\times(-\ga^0\ga^5)\times P_{\nu}^{*}e^{\nu}= -m d^{00} \ga^0\ga^5 P_{\nu}^{*}e^{\nu}.
\label{doo}
\end{equation}
Observe that both of them break $C$, $P$, $PT$ and $CT$ \cite{tables, pickering}. It is remarkable to say that 
the study of this kind of terms, with EFWT 
considerations, depend on the correct choice of the involution operator, such that 
relation (\ref{atrfg}) is contemplated.

The general form of the involution operator \cite{nikitin, violeta}  
has the following structure

\begin{equation}
\hat{J}=M\times\hat{F}\,, 
\label{struc}
\end{equation}
where M and $\hat{F}$ are operators. They act on the matrices and functions space, respectively.
In particular, the choice $M=i\ga^5\ga^0$ and $\hat{F}=\hat{1}$ corresponds  to the usual operator used in previous works \cite{bruno2, bmb}. However, as already mentioned above, 
the new terms in the table do not satisfy the anti-commutation relation (\ref{atrfg}), for such a choice. The main point here is the following: 
the choice of an appropriated involution operator, for a 
specific term of the table, involves the knowledge of exactly what symmetry is being broken (for each term of the table).  

An interesting case is the vectorial 
part of the torsion field, $b_l$. As one can check \cite{tables, pickering}, this term breaks $T$, $CT$, $PT$ and $CPT$.
In the Hamiltonian, the torsion field is founded by the product  
of line $0$ by row $6$. It gives $b_l\ga^0 \ga^5\ga^l$. It was showed that  $M=i\ga^5\ga^0$ and $\hat{F}=T$ represents a specific choice for the involution operator, such that the anti-commutation relation is obeyed \cite{bmb}. However, it is not the only possible choice. In particular $\hat{F}=CT$, $\hat{F}=PT$ and $\hat{F}=CPT$, would work equally. 
 
In order to perform the EFWT for all the terms in the table above, we present in the next section, a proposal of a new involution operator which anti-commutes with all the terms of the table.


\section{The involution operator, an appropriated choice}

\hspace{0.65cm}We begin with an appropriated representation of the Hamiltonian with CPT-Lorentz breaking terms. 

\begin{equation}
{\cal H}=\phi_{1}^{A}\,H_{A B}\,\phi_{2}^{B} \,.
\label{nham}
\end{equation}
Throughout this paper, the quantities with Latin indexes $A$ and $B$ are only associated with possible positions in the table. 
We would like to emphasize that these indexes are not space-time indexes. The possible values for $A$ are running horizontally in the table, from 0 to 8. In the case of $B$, the possible values 
are running vertically, from 0 to 9. In addition, the quantities $\phi_{1}^{A}$ and $\phi_{2}^{B}$ are the fields that appear in the top row and in the left column of the table, respectively, and the quantity $H_{AB}$ represents the terms contained in the cells of the table.

As shall be better understood in the next section, 
the EFWT works if, and only if, one can write the Hamiltonian on the form of Eq. (\ref{nham}). 
It may seem cumbersome, at a first sight, but it is not. 
Let us consider an example. 
The choice $A=0$ and $B=6$ (first column and 
the seventh row, respectively) leads us to $\phi_{1}^{0}=m+P_{\nu}^{*}e^{\nu}$, $\phi_{2}^{6}=d_{00}$ and $H_{0,6}=-\ga^0\ga^5$\,. 
It gives exactly the two terms described in Eq. (\ref{doo}). 

Taking into account these considerations, we present, as a next step, an 
involution operator which anti commutes with the complete set of term of the Hamiltonian (\ref{nham}).
\begin{equation}
\hat{J}=\big(i\ga^5\ga^0\big)\,\times\, 
\big(C^{{\cal O'}_{AB}}P^{{\cal O''}_{AB}}T^{{\cal O'''}_{AB}}\big)^{\theta_{IK}}\,, \label{jota}
\end{equation}
where $C$, $P$ and $T$ are the known charge, parity and time operators, respectively \cite{tables, pickering}. 
Observe that Eq. (\ref{jota}) obeys the structure of Eq. (\ref{struc}), with the following $M$ and $\hat{F}$ choice
\begin{equation}
M=i\ga^5\ga^0 \hspace{1cm} \mbox{and} \hspace{1cm} 
\hat{F}=\big(C^{{\cal O'}_{AB}}P^{{\cal O''}_{AB}}T^{{\cal O'''}_{AB}}\big)^{\theta_{IK}}
\label{opNOVO}
\end{equation}
where we define,
\begin{equation}
I=A-5 \hspace{1cm} \mbox{and} \hspace{1cm} K=B-6 \label{ij}\,.
\end{equation}

The quantity $\theta_{IK}$ is defined in order to assume the values $0$ or $1$. If $I\times K>0$, $\theta_{IK}=0$. 
On the other hand, if $I\times K<0$, 
$\theta_{IK}=1$. Actually, the product between $I$ and $K$ tells us if we are dealing with the new or old terms 
of the table. 
The quantities ${{\cal O}_{AB}}$ also assume the $0$ or $1$ values. They are 
determined by the previous knowledge of which symmetry is being broken. 

Let us consider an example, by setting $A=6$ and $B=0$. Then, $\phi_{1}^{6}=b_l+ P^{*}_{\nu}d^{ln}$, 
$\phi_{2}^{0}=\ga^0$ and $H_{6,0}=\ga^5\ga^l$. 
According to Eq. (\ref{nham}), the Hamiltonian for this case, is given by
\begin{equation}
{\cal H}=\ga^0\ga^5\ga^l\big(b_l +P^{*}_{\nu}d^{ln}\big)\,.\label{gg}
\end{equation}

The next step is the choice of the ${{\cal O}_{AB}}$ quantities. In Refs. \cite{tables, pickering}, there is a table 
with the properties of operators for Lorentz violation in QED. According to this table, one can 
consider that ${{\cal O''}_{AB}}=0$ and ${{\cal O'}_{AB}}={{\cal O'''}_{AB}}=1$. Observe that 
from the Eq. (\ref{ij}), $I=1$, $K=-6$ and $I\times K=-6$, for this reason we have $\theta_{6,0}=1$.
With these considerations, the corresponding involution operator is 
\begin{equation}
\hat{J}=i\ga^5\ga^0\, PT\,. \label{jct}
\end{equation}
As one can check, the anti-commutation relation is obeyed, when the the quantities ${\cal H}$ and $\hat{J}$ 
are described by the relations (\ref{gg}) and (\ref{jct}), respectively.

One can see that for the old terms of table, the product between $I$ and $K$ is always 
positive and the quantity $\theta$ in Eq. (\ref{jota}) is equal to zero. Consequently, in what 
concerns the old part of the table, we shall have, as expected, $M=i\ga^5\ga^0$ and $\hat{F}=\hat{1}$.


\section{Exact transformation with CPT}

\hspace{0.65cm} We present in this section, the EFWT of the Hamiltonian for a free spin-$1/2$ Dirac fermion $\Psi$ of 
mass m in the standard-model extension \cite{colladay, kostebase}. Let us begin with the following Hamiltonian 

\begin{eqnarray}
{\cal H}&=&m\big( \ga^0-\ga^0c_{00}-e_0-d_{j0}\ga^{5}\ga^j+\frac{1}{2}g_{ik0}\si^{ik} \big)\nonumber\\
&+&{P}^k\big(-\al_{k}+2d_{0k}\ga^{5}-c_{jk}\al^{j}+c_{00}\al_{k}+if_k\ga^5\ga^0 
- 2g_{0jk}\ga^0\si^{0j}-2c_{0k}+d_{jk}\ga^5\al^j\nonumber \\
&-&d_{00}\ga^5\al_k -e_k\ga^0 +\frac{1}{2}\ga^0\si^{ij}g_{ijk}-ig_{i00}\ga_k\al^i\big)\nonumber \\
&+&a_j\al^j-b_0\ga^5+iH_{0j}\ga^j-b_j\ga^5\al^j-\frac{1}{2}\ga^0\si^{ij}H_{ij}\,.
\label{uma}
\end{eqnarray}
This Hamiltonian can be constructed directly from the table presented in the last section. However, it is 
not the most complete Hamiltonian that one can extract from the table. The main point of this work is the development of the operator described in (\ref{opNOVO}). As it is been used 
for the first time, it is worthwhile to deal with a Hamiltonian which the diagonalized result 
we could know at least the qualitative result. On the other hand, it would be very interesting from the physical point of view if the the new EFWT generates unexpected terms in comparison with the usual transformation for the same action. We decide to pick just the 
terms represented in Eq.(\ref{uma}) because in \cite{kostebase} the authors perform the usual FWT, taking into account this Hamiltonian. Performing the transformation for it we could 
validate our algorithm and also search for physical quantities mixed in a new form. The transformed Hamiltonian (with usual FWT) is the following\footnote{Now and so on, 
we denote transformed quantities by using the "tr" index.} 
\begin{equation}
{\tilde{H}^{tr}} = \be m + \frac{1}{2m}\Big\{(1+\tilde{A})\Big[\Big(\de_{ij}
+{\tilde{B}}_{ij}\Big)\bar{P}^i+{\tilde{C}}_{j}\Big]^2+{\tilde{D}}\Big\}\,,
\end{equation}
where
\begin{eqnarray}
{\tilde{A}} &=& -2c_{00}\ga^{0} \nonumber \\
{\tilde{B}_{ij}} &=& \frac{1}{2}\Big[4\Big(d_{0i}+d_{i0}\Big)\ga^{5}\ga^{j} - 4c_{ij}\ga^{0} + 4{\epsilon^{l}}_{mj}\Big(g_{l0i}+g_{li0}\Big)\ga^{5}\ga^{0}\ga^{m}\Big] \nonumber\\
{\tilde{C}_{j}} &=& \frac{1}{2}\Big[-4m\Big(c_{0j}+c_{j0}\Big) + 4md_{ij}\ga^{5}\ga^{0}\ga^{i} - 4md_{00}\ga^{5}\ga^{0}\ga^{j} - 4me_{j}\ga^{0} + 2m{\epsilon^{kl}}_{m}g_{klj}\ga^{5}\ga^{m} \nonumber\\
&-& 4m{\epsilon^{ij}}_{l}g_{i00}\ga^{5}\ga^{l} + 4a_{j}\ga^{0} - 4b_{0}\ga^{5}\ga^{j} + 4{\epsilon^{jk}}_{l}H_{0k}\ga^{5}\ga^{0}\ga^{l}\Big] \nonumber\\
{\tilde{D}} &=& -2m^{2}c_{00}\ga^{0} - 2m^{2}e_{0} - 2m^{2}d_{j0}\ga^{5}\ga^{j} - m^{2}{\epsilon^{ik}}_{l}g_{ik0}\ga^{5}\ga^{0}\ga^{l} \nonumber\\
&+& 2ma_{0} - 2mb_{j}\ga^{5}\ga^{0}\ga^{j} + m{\epsilon^{ij}}_{l}H_{ij}\ga^{5}\ga^{l}
\label{finalKost}
\end{eqnarray}
Besides EWFT is 
more economic in algebra, it presents more detailed information 
with respect to the non-relativistic approximation \cite{FW, mckellar, silenko4, silenko5}. 

As a first step to perform the EFWT, we calculate the squared Hamiltonian ${\cal H}^2$. 
In order to simplify the the algebra, we shall write this quantity as 
\begin{equation}
{\cal H}^2=m^2\Big(1+\frac{\bar{{\cal H}}^2}{m^2}\Big)\label{kim} \,,
\end{equation}
where $\bar{{\cal H}}^2$ is given by
\begin{equation}
\bar{{\cal H}}^2= (1+\bar{A})[(\de_{ij}+\bar{B}_{ij})\bar{P}^i+\bar{C}_j]^2+\bar{D}\,.
\end{equation}
The quantities $\bar{A}$, $\bar{B}_{ij}$, $\bar{C}_j$ and $\bar{D}$ are 
written in the form
\begin{eqnarray}
\bar{A}&=& -2c_{00}-d_{00}\ga^{5}+2ig_{i00}\ga^0\al^i \,, \nonumber \\
\bar{B}_{ij}&=& \frac{1}{2}\Big[-8\,d_{0i}\ga^5\al_j-4c_{ij}+8\,g_{0li}\ga^0\epsilon^{jlm}\Sigma_m
+ 8\,c_{0i}\al_j+4d_{ij}\ga^5 + 4g_{lmi}\epsilon^{lmj}\ga^0\ga^5 \nonumber \\
&+& 4ig_{ilj}\ga^{0}\ga^5\Sigma^{l} +4ig_{i00}\ga^0\al_j \Big] \,, \nonumber \\
\bar{C}_j&=&\frac{1}{2}\Big[-8m\ga^0c_{0j}+4md_{ij}\ga^0\ga^5\al^i-4md_{00}\ga^0\ga^5\al_j 
-4me_j +2mg_{klj}\si^{kl}-4img_{j00}\nonumber \\
&-&4mg_{i00}\epsilon^{ijl}\Sigma_l +4me_0\al_j+4imd_{k0}\ga^0\ga^5\epsilon^{jkl}\Sigma_{l} 
-2m g_{il0}\epsilon^{ilj}\ga^{5} +4a_j + 4b_0\ga^5\al_j  \nonumber \\
&-& 4H_{0k}\ga^0\epsilon^{jkl}\Sigma_l -4a_0\al_j -4b_j\ga^5-4H_{kl}\epsilon^{klj}\ga^0\ga^5
+4iH_{lj}\ga^0\ga^5\Sigma^{l}\Big]\,,\nonumber\\
\bar{D}&=&-2m^2c_{00}-2m^2\ga^0e_0+2m^2d_{j0}\ga^5\al^j + m^2\ga^0\si^{ik}g_{ik0}
+2m\ga^0a_0-2m\ga^0\ga^5\al^jb_j \nonumber\\
&-& m\si^{ij}H_{ij}+\Big(1-2c_{00}+2d_{00}\ga^5-2ig_{i00}\ga^0\al^i\Big)
\frac{i\hbar e}{mc}\Sigma_k B^k\,.
\end{eqnarray} 

There are, in the last equation, even and odd terms. In the FW context, even and odd operators are written as 
\begin{equation}
M_{(EVEN)}=\frac{1}{2}(M+\ga^0 M\ga^0)\hspace{1cm} \mbox{and}\hspace{1cm}
M_{(ODD)}=\frac{1}{2}(M-\ga^0 M\ga^0)\,.
\end{equation}
In the situation when there are many odd terms, one must take into account the following relation \cite{bmb} 
\begin{equation}  
{\cal H}^{tr}=\hat{J}\frac{1}{2}(\sqrt{{\cal H}^2}-\ga^0 \sqrt{{\cal H}^2}\ga^0) 
+\ga^0\frac{1}{2}(\sqrt{{\cal H}^2}+\ga^0 \sqrt{{\cal H}^2}\ga^0)\,,\label{hamtr}
\end{equation}
where $\hat{J}$ is given by Eq. (\ref{jota}). The transformed Hamiltonian is denoted by ${\cal H}^{tr}$ which presents only even terms. For this reason, 
${\cal H}^{tr}$ does not mix spinor components. 
Naturally, the calculation of $\sqrt{{\cal H}^2}$ should be performed and the result 
inserted in the Eq. (\ref{hamtr}). Let us consider that $m^2 \gg \bar{{\cal H}}^2$ in the Eq. (\ref{kim}), such that 
\begin{equation}
\sqrt{{\cal H}}=m\Big(1+\frac{\bar{{\cal H}}^2}{2m^2}\Big)\,.
\end{equation}
After some algebra, the transformed Hamiltonian is given by
\begin{equation}
 {\cal H}^{tr}=\ga^0 m+\frac{1}{2m} \Big\{ (1+A^{tr})\Big[(\de_{ij}
+B_{ij}^{tr})\bar{P}^{i}+C_{j}^{tr}\Big]+D^{tr}\Big\}\,,
\end{equation}
where 
\begin{eqnarray}
A^{tr}&=&-2\ga^0c_{00}-2i\ga^{0}d_{00}+2g_{i00}\,,\Sigma^i\nonumber\\
B_{ij}^{tr}&=&\frac{1}{2}\Big[8d_{0i}\ga^0\Sigma_{j}-4\ga_{0}c_{ij} 
-8g_{0li}\epsilon^{jlm}\Sigma_{m}-8ic_{0i}\ga^0\Sigma_{j}+4i\ga^0d_{ij}\nonumber \\
&+&4ig_{lmi}\epsilon^{lmj} -4g_{ilj}\Sigma^{l} +4g_{i00}\Sigma^{i}\Big]\,,\nonumber \\
C_{j}^{tr}&=&\frac{1}{2}\Big[8mc_{0j}+4md_{ij}\Sigma^{i}-4md_{00}\Sigma_{j}-4m\ga^{0}e_{j}
+2mg_{klj}\ga^0\epsilon^{klm}\Si_{m}-4img_{j00}\ga^0 \nonumber \\
&-&4mg_{i00}\ga^0\epsilon^{ijl}\Si_{l} -4ime_{0}\ga^{0}\Si_j 
-4md_{k0}\epsilon^{jkl}\Si_l-2img_{il0}\epsilon^{ilj}\ga^0+4\ga^0a_j \nonumber \\
&-&4b_0\ga^0\Si_j 
+4H_{0k}\epsilon^{jkl}\Si_l +4ia_0\ga^0\Si_j -4ib_j\ga^0 -4i\epsilon^{klj}H_{kl} 
-4H_{lj}\Si^l\Big]\,, \nonumber \\
D^{tr}&=& -2m^2\ga^0c_{00} +2m^2e_0 -2m^2d_{j0}\ga^0\Si^j-m^2g_{ik0}\epsilon^{ikl}\Si_l -2ma_0
-2mb_j\Si^j\nonumber \\
&-&m\ga^0\epsilon^{ijl}H_{ij}\Si_l +\ga^0\big( 1+2c_{00}+2id_{00}-2g_{i00}\ga^0\Si^i\big)
\frac{i\hbar e}{mc}\Sigma_k B^k\,.
\label{final2016}
\end{eqnarray}
We have considered Eq. (\ref{uma}) as the starting point,
in order to obtain the transformed Hamiltonian (\ref{final2016}). 
It is possible to see that there are nine new terms in (\ref{final2016}) when compared to  (\ref{finalKost}). The new terms are one in the quantity $\bar{A}$ related to the coefficient $d_{00}$; two in $\bar{B}_{ij}$ related to the coefficients $c_{0i}$ and $d_{ij}$; four in $\bar{C}_j$ related to the coefficients $a_{0}$, $b_{j}$, $e_{0}$ and $g_{il0}$; and two terms in $\bar{D}$ related to the coefficients $c_{00}$ and $d_{00}$ with the magnetic field.
Nevertheless, the exact process has some advantages when compared to the usual one, as commented above. For instance, the new terms that appear in $D$, are relevant when the bound state of the theory is considered. 


\section{Bound State of the theory}

\hspace{0.65cm} The determination of which kind of experimental tests, like Penning trap, 
Clock comparison, torsion pendulum, among others (see \cite{blumblum1, lane, blumblum2, blumblum3, blumblum4}, 
and references cited there in.),  has a significant 
relevance in the scope of the standard model extension (SME) \cite{colladay}. In order to determine 
the kind of experimental test that should be performed, considering the 
CPT-Lorentz violation terms, presented in the Dirac equation, one should derive 
the bound state of the theory. In Ref. \cite{tables}, the authors present a table with a set 
of many possible bound states. It is expected that the bound associated 
with transformed Hamiltonian (\ref{final2016}), 
could be found in such a table. 
Hence, with the knowledge of the bound, together 
its magnitude and the original Hamiltonian, 
one can determine the kind of appropriated experimental test should be performed 
(See \cite{colladay, koste1, koste2, perry} for a theoretical framework about CPT-Lorentz breaking tests). 

In this section, we derive the bound state of the Hamiltonian (\ref{final2016}). 
Let us begin by taking into account the two components spinor 
\begin{equation}
\psi=\left(
\begin{array}{ccc}
\phi \\
\chi  \\
\end{array}
\right) \,\exp^{-imt}\,.
\end{equation}
From this point one can write, after some algebra,  the Dirac equation in the 
Schr\" odinger form $i\partial_{t}\psi={\cal H}\psi$. With these considerations, 
the Hamiltonian to $\phi$ is written as
\begin{equation}
{\cal H}= \frac{1}{2m}\big\{(1+A)[\big(\de_{ij}
+B_{ij}\big)\bar{P}^i+C_{j}]^2+D\big\}\,,
\end{equation}
where
\begin{eqnarray}
A &=& -2c_{00}-2id_{00}+2g_{i00}\sigma^{i}\nonumber\\
B_{ij} &=& 4d_{0i}\sigma_{j}-2c_{ij} 
-4g_{0li}\epsilon^{jlm}\sigma_{m}-4ic_{0i}\sigma_{j}+2id_{ij}\nonumber \\
&+&2ig_{lmi}\epsilon^{lmj} -2g_{ilj}\sigma^{l} +2g_{i00}\sigma^{i}\,,\nonumber \\
C_{j} &=& 4mc_{0j}+2md_{ij}\sigma^{i}-2md_{00}\sigma_{j}-2me_{j}
+mg_{klj}\epsilon^{klm}\sigma_{m}-2img_{j00} \nonumber \\
&-&2mg_{i00}\epsilon^{ijl}\sigma_{l} -2ime_{0}\sigma_j 
-2md_{k0}\epsilon^{jkl}\sigma_{l}-img_{il0}\epsilon^{ilj}+2a_j \nonumber \\
&-&2b_0\sigma_{j} 
+2H_{0k}\epsilon^{jkl}\sigma_{l} +2ia_0\sigma_{j} -2ib_j -2i\epsilon^{klj}H_{kl} 
-2H_{lj}\sigma^{l}\,, \nonumber \\
D &=& -2m^{2}c_{00} +2m^{2}e_0 -2m^{2}d_{j0}\sigma^{j}-m^2g_{ik0}\epsilon^{ikl}\sigma_{l} -2ma_0
-2mb_j\sigma^{j}\nonumber \\
&-&m\epsilon^{ijl}H_{ij}\sigma_{l} +\Big[ 1+2c_{00}+2id_{00}-2g_{i00}\sigma^{i}\Big]
\frac{i\hbar e}{mc}\sigma_{k} B^k\,.
\label{eqSpin}
\end{eqnarray}
The bound state of the Hamiltonian (\ref{final2016}) can be calculated 
by taking into account the Lorentz violating potential $V$, which corresponds 
to the term $D$, in the last equation. Actually, this potential obeys the following relation \cite{lane}
 \begin{equation}
V=-\tilde{b}_{j}\si^{j}\,,
\end{equation}
where $\si$ represents the spin matrices. From this point, one can calculate 
the bound state of the theory:
\begin{equation}
\tilde{b}_j = b_j +\frac{1}{2}\epsilon^{lmj}H_{lm} +md_{j0} 
+\frac{1}{2}m\epsilon^{lmj}g_{lm0} -\Big[1+2c_{00}+2id_{00}
-2g_{i00}\sigma^{i}\Big]\frac{i\hbar e}{2 m^2 c}B_{j}\,.
\label{bound}
\end{equation}
As it was expected, this bound state is a specific
combination of two parts related to the SME coefficients. The first part includes the coefficients $b_{j}$, $H_{lm}$, $d_{j0}$ and $g_{lm0}$, and the bound is based on atomic clock and other non-relativistic experiments \cite{altschul} that can involve maser/magnetometer (see, for example, table VII in \cite{tables}). In the second part, there is  the presence of magnetic field which can be a remarkable and very important result from the experimental point of view. As it is known, the external fields in the Eq. (\ref{bound}), are very weak. 
However, the modulus of $\vec{B}$ may be sufficiently high, in order to compensate the weakness of 
the interactions $c_{00}$, $d_{00}$ and $g_{i00}$. In another words, with a strong enough magnetic field, one can 
have indications, in principle, of the kind of motion generated by the external field commented above. 
It is, an indirect way of performing measurements of such a weak external fields. 



\section{Conclusions and discussions}
\label{Con}

\hspace{0.65cm}

The exact Foldy-Wouthuysen transformation was performed in the context of Dirac 
field interacting with many possible external fields associated with CPT-Lorentz 
violation. 

The first result of the work is written in the form of a table, representing 
the Hamiltonian with the complete set CPT-Lorentz violating terms, in the Dirac equation.
In such table, the terms highlighted in boldface do not anti-commute with the usual 
involution operator (\ref{oldj}).

Another result of the work is the appropriated 
involution operator, given by Eq. (\ref{jota}), such that the anti-commutation 
relation with the Hamiltonian of the problem is achieved. Actually, 
Eq. (\ref{jota}), introduces a new possibility 
of performing EFWT. From now on, a large class of Hamiltonians admit the exact transformation, 
since involution operator (\ref{jota}) is used. 
 
In section 4, the usual EFWT algorithm 
was applied to the initial Hamiltonian and the exact transformation was performed. 
As it was expected, the EWFT approach presents a transformed Hamiltonian (\ref{finalKost}) with 
additional terms, when compared 
to the Hamiltonian (\ref{final2016}), where the usual FWT is used. 

In the last section we derive the bound state of the theory, given by Eq. (\ref{bound}). 
It worth mentioning that the possibility of the weakness of 
CPT-Lorentz terms to be compensated by the presence of a strong magnetic field. 
Thus, one can understand the particle behavior due to the interactions with external field, 
it gives the possibility to measure the external fields in indirect way.


\section*{Acknowledgments}

\hspace{0.65cm} The authors wish to thank Prof. Ilya L. Shapiro for the initial discussions about the problem.
BG is grateful to Funda\c c\~ ao de Amparo a Pesquisa de Minas Gerais (FAPEMIG) and 
Funda\c c\~ ao Nacional de Desenvolvimento da Educa\c c\~ ao (FNDE) for financial support. 
MDJ is grateful to the Programa de Bolsas de P\' os-Gradua\c{c}\~ ao da UFJF (PBPG-UFJF).


\renewcommand{\baselinestretch}{0.9}

\end{document}